%% file: root.tex
\documentclass[letterpaper, 10 pt, conference]{ieeeconf}
\IEEEoverridecommandlockouts    
\usepackage{graphics} 
\usepackage{epsfig} 
\usepackage{times} 
\usepackage{amsmath} 
\usepackage{amssymb}  

\usepackage{subfigure}
\usepackage{mathtools}
\usepackage{color}
\usepackage{cite}
\usepackage{comment}
\usepackage{balance}
\usepackage[table]{xcolor}
\usepackage{multirow}
\usepackage{hyperref}
\hypersetup{
    colorlinks=true,
    linkcolor=black,
    citecolor=green,
    urlcolor=blue,
}
\usepackage{blindtext}
\usepackage{bm}
\usepackage{algorithm}
\usepackage{algpseudocode}
\usepackage[official]{eurosym}

\makeatletter
\let\pragma@iinput=\@iinput
\def\@iinput#1{\xdef\@pragmafile{#1}\pragma@iinput{#1}}
\def\@pragmafile{default}
\def\pragmaonce{%
   \csname pragma@\@pragmafile\endcsname
   \global\expandafter\let \csname pragma@\@pragmafile\endcsname = \endinput
}
\makeatother

\algrenewcommand\algorithmicindent{1.0em}
\algnewcommand\algorithmicswitch{\textbf{switch}}
\algnewcommand\algorithmiccase{\textbf{case}}
\algnewcommand\algorithmicassert{\texttt{assert}}
\algnewcommand\Assert[1]{\State \algorithmicassert(#1)}%
\algdef{SE}[SWITCH]{Switch}{EndSwitch}[1]{\algorithmicswitch\ #1\ \algorithmicdo}{\algorithmicend\ \algorithmicswitch}%
\algdef{SE}[CASE]{Case}{EndCase}[1]{\algorithmiccase\ #1}{\algorithmicend\ \algorithmiccase}%
\algtext*{EndSwitch}%
\algtext*{EndCase}%

\makeatletter
\newcommand{\algmargin}{\the\ALG@thistlm}
\makeatother
\newlength{\forwidth}
\algdef{SE}[parFOR]{parFor}{EndparFor}[1]
  {\parbox[t]{\dimexpr\linewidth-\algmargin}{%
     \hangindent\forwidth\strut\algorithmicfor\ #1\ \algorithmicdo\strut}}{\algorithmicend\ \algorithmicfor}%
\newlength{\forif}
\algnewcommand{\parState}[1]{\State%
  \parbox[t]{\dimexpr\linewidth-\algmargin}{\strut #1\strut}}

\newtheorem{definition}{Definition}[section]

\title{\LARGE \bf
Ensuring Safety at Intelligent Intersections: \\ Temporal Logic Meets Reachability Analysis
}

\author{Kaj Munhoz Arfvidsson$^*$, Frank J. Jiang$^*$, Karl H. Johansson, Jonas Mårtensson%
\thanks{
    $^*$Indicates equal contribution}%
\thanks{
        This work was partially supported by the Wallenberg Artificial Intelligence, Autonomous Systems, and Software Program (WASP) funded by the Knut and Alice Wallenberg Foundation. It was also partially supported by the Swedish Research Council, Swedish Research Council Distinguished Professor Grant 2017-01078, the Knut and Alice Wallenberg Foundation Wallenberg Scholar Grant, and the Swedish Innovation agency (Vinnova), under grant 2021-02555 Future 5G Ride, within the Strategic Vehicle Research and Innovation program (FFI).
        }
\thanks{
    All authors are with the Division of Decision and Control Systems, EECS, KTH Royal Institute of Technology, Malvinas v{\"a}g 10, 10044 Stockholm, Sweden {\tt\small \{kajarf, frankji, kallej, jonas1\}@kth.se}. They are also affiliated with the Integrated Transport Research Lab and Digital Futures.}%
}

\begin{document}

\maketitle
\thispagestyle{empty}
\pagestyle{empty}

\begin{abstract}

In this work, we propose an approach for ensuring the safety of vehicles passing through an intelligent intersection. There are many proposals for the design of intelligent intersections that introduce central decision-makers to intersections for enhancing the efficiency and safety of the vehicles. To guarantee the safety of such designs, we develop a safety framework for intersections based on temporal logic and reachability analysis. We start by specifying the required behavior for all the vehicles that need to pass through the intersection as linear temporal logic formula. Then, using temporal logic trees, we break down the linear temporal logic specification into a series of Hamilton-Jacobi reachability analyses in an automated fashion. By successfully constructing the temporal logic tree through reachability analysis, we verify the feasibility of the intersection specification. By taking this approach, we enable a safety framework that is able to automatically provide safety guarantees on new intersection behavior specifications. To evaluate our approach, we implement the framework on a simulated T-intersection, where we show that we can check and guarantee the safety of vehicles with potentially conflicting paths.

\end{abstract}


\input{src/1_intro}

\input{src/2_prelim}

\input{src/3_mot_ex}

\input{src/4_method}

\input{src/5_experiments}

\input{src/6_conclusion}

\balance


\bibliographystyle{IEEEtran}
\bibliography{refs}




\end{document}

%% file: src/1_intro.tex
\section{Introduction}\label{sec:intro}
In recent years, the need for intelligent intersections in the transportation network has become increasingly evident. Since intersections are often both inefficient and dangerous~\cite{Grembek2018}, there is a significant amount of work that has gone into proposing updates to traditional intersection management techniques that involve higher levels of autonomy and vehicle-to-infrastructure (V2I) communication, e.g.~\cite{Dresner2004, Namazi2019}.
While there is a significant emphasis on safety for the design of intelligent intersections, there is still little consensus around how we should provide safety guarantees that are flexible to possible changes in the intersection specification. One of the core challenges is the difficulty in computing the maximal controlled invariant sets for intersections in a general, computationally-tractable way, since finding the exact solution is an NP-complete problem~\cite{Colombo2012}. To address this, several approaches propose approximate solutions
to the problems where the maximal controlled invariant set is conservatively approximated and leverage assumptions made about how the vehicles will pass through the intersection~\cite{Colombo2012, Chen2022, Kowshik2011}. Alternatively, some take a probabilistic approach and provide lower bounds on vehicle collision probabilities in intelligent intersections~\cite{Thunberg2021}.
While these approaches do provide safety guarantees, they are built upon specific intersection traffic rules. Due to the diverse and evolving requirements of traffic passing through intersections, there is interest to further investigate approaches that are able to provide more flexible safety guarantees that easily adapt to updates to changing intersection traffic rules.

\begin{figure}
    \centering
    \vspace{0.25cm}
    \includegraphics[width=0.4\textwidth]{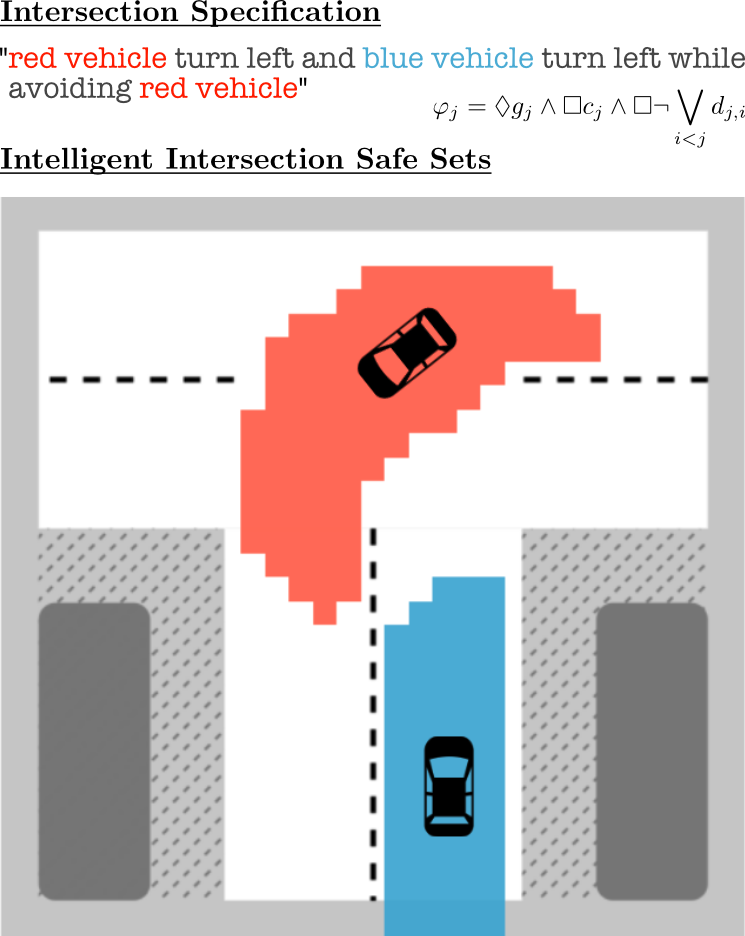}
    \caption{T-Intersection example with two vehicles passing through and the safe sets computed for the intersection specification.}
    \label{fig:example}
\end{figure}

For more flexible safety guarantees in intelligent transportation systems, researchers have recently proposed several approaches based on the formalization of traffic rules. Used in the specification and verification of various types of complex systems~\cite{Baier2008}, temporal logic offers a compelling approach for formalizing requirements on systems in a way that is both flexible and approachable to human designers. For example, \cite{Maierhofer2022}~shows that they can formalize current German intersection traffic rules using metric temporal logic. Specifically for designing intersection management, \cite{Saraoglu2022}~use linear temporal logic to specify and verify the safety of an intersection management algorithm. While they do not use temporal logic, \cite{Haydon2023}~similarly develop formal specifications for intersection management by formalizing the responsibility-sensitive safety model~\cite{Shalev-Shwartz_Shammah_Shashua_2017} using Hoare Logic that can be used for discovering conditions that guarantee safety of the intersection. In this work, we show how to take intersection rules that are formalized in linear temporal logic and leverage temporal logic trees~\cite{Gao_Abate_Jiang_Giacobbe_Xie_Johansson_2022} to directly use Hamilton-Jacobi (HJ) reachability analysis to verify the feasibility of the intersection rules.

The main contribution of this paper is a safety framework that verifies the feasibility of a multi-vehicle specification in an intelligent intersection. Specifically, the contributions of the paper can be summarized as follows:
\begin{enumerate}
    \item we present a linear temporal logic-based sequential path planning approach for intelligent intersections,
    \item we detail the construction of temporal logic trees for verifying the feasibility of the sequential path planning,
    \item we evaluate the approach by verifying the safe crossing of vehicles through a T-intersection. The code used for the evaluation is publicly available\footnote{https://github.com/kaarmu/safe\_intersections}.
\end{enumerate}
Taking inspiration from the sequential path planning approaches developed for aerial vehicles in~\cite{Chen_Bansal_Tanabe_Tomlin_2017, Bansal_Chen_Tanabe_Tomlin_2021}, the contributed approach starts with formalizing an intersection specification with linear temporal logic formulae. Then, by using temporal logic trees for the verification of these formulae, we develop an approach that results in both safety guarantees for the intelligent intersection and is also able to automatically handle any changes to the specifications. Through the use of HJ reachability analysis, the approach is also able to handle the nonlinearities of road vehicle models and the complex environments of intersections. Moreover, in our numerical evaluation, we find preliminary indications that, by using modern software libraries, the online verification computation time is potentially fast enough to be implemented on real intelligent management systems.

The remainder of the paper is organized as follows. In Section~\ref{sec:prelim}, we provide the necessary preliminary material for the presented approach and use a motivating example to state the problem addressed in this work.
In Section~\ref{sec:method}, we describe our approach to verifying the feasibility and safety of temporal logic specifications for intelligent intersections. In Section~\ref{sec:exp}, we numerically evaluate the approach on a three vehicle T-intersection scenario. In Section~\ref{sec:conc}, we conclude the paper with a discussion about our work and future directions.

%% file: src/2_prelim.tex
\section{Problem Formulation}\label{sec:prelim}

\input{src/commands}

In this section, we introduce the necessary preliminary material for our approach. We focus on introducing the notation and definitions that are key for the application of temporal logic and reachability analysis to ensuring safety at intelligent intersections. Finally, we state the specific problem addressed in this work.

\subsection{Multi-Vehicle Model}
In this section, we define the multi-vehicle model we use for ensuring the safety of intelligent intersections. We start by defining the vehicle model for a single vehicle and then collect the single vehicle models into a collective multi-vehicle model.

For a single vehicle $i$, let $z_i = [x_i, y_i, \theta_i, \delta_i, v_i]^\top$ be the state, where $x_i$, $y_i$, $\theta_i$, $\delta_i$, and $v_i$ are the vehicle's x-position, y-position, heading angle, steering angle, and velocity, respectively. Then, let $u_i = [s_i, a_i]^\top$ be the input, where $s$ and $a$ are the steering rate and acceleration inputs into the vehicle, respectively. Explicitly, we model the dynamics with
\begin{equation}\label{eq:single_vehicle}
    \dot{z}_i = f_i(z_i) + g_i(z_i)u_i,
\end{equation}
where
\begin{equation}
    f_i(z_i) =
    \left[\begin{matrix}
        v_i \cos\theta_i \\
        v_i \sin\theta_i \\
        \frac{v_i \tan \delta_i}{L_i}\\
        0 \\
        0
    \end{matrix}\right], \ 
    g_i(z_i) = 
    \left[\begin{matrix}
        0 & 0\\
        0 & 0\\
        0 & 0\\
        1 & 0 \\
        0 & 1
    \end{matrix}\right].
    \nonumber
\end{equation}
$L_i$ is the wheel-base length of the vehicle. Here, $z_i\in\mathbb{R}^5$ and $u_i \in \mathbb{U} \subset \mathbb{R}^2$, where $\mathbb{U}$ is the full set of physically feasible steering rates and accelerations/decelerations we assume vehicles passing through the intersection can implement. To analyze the possible decisions vehicles could make while passing through the intersection, we let $u_i(\cdot) \in U$ be vehicle $i$'s physically feasible control policy, where $U$ is the function space containing all physically feasible vehicle control policies. We also denote a trajectory of vehicle $i$ with $\zeta_i(\cdot; z_{i,0}, t_0, u_i(\cdot))$, which is the trajectory starting from initial state $z_{i, 0} = \zeta_i(t_0; z_{i,0}, t_0, u_i(\cdot))$ under control policy $u_i(\cdot)$. For simplicity, we sometimes describe the trajectory of vehicle $i$ with $\zeta_i(\cdot)$.

For analyzing the behavior of multiple vehicles in an intersection, we also define a multi-vehicle model. For an intersection with $N$ vehicles, let the full multi-vehicle state and control input be $z = [z_1, z_2, \ldots, z_N]^\top$ and $u = [u_1, u_2, \ldots, u_N]^\top$, respectively. Then, we can write the full multi-vehicle dynamics as the following:
\begin{equation}\label{eq:multi_vehicle}
    \dot{z} = f(z) + g(z)u,
\end{equation}
Finally, we write the collective trajectory of the multi-vehicle system as $\zeta(\cdot; z_0, t_0, u(\cdot))$, which we will also sometimes write as $\zeta(\cdot)$ for simplicity.

For the remainder of this paper, we will work with time-state sets of of the multi-vehicle system. First, denote the full state space of the multi-vehicle system as $\mathbb S$. We denote time-state sets for the multi-vehicle system as $\mathcal S \subseteq \mathbb S \times \mathbb R$. Then, for retrieving state sets at particular times we define the time-state set map $\Omega: \mathbb R \rightarrow \mathbb S$. An $\mathcal S$ has a corresponding $\Omega_{\mathcal{S}}$ where
\begin{equation}
    \mathcal S = \bigcup_{t \in \mathbb R} \{ (z,t) \mid z\in\Omega_{\mathcal S}(t) \}.
\end{equation}
If, however, $\{ (z, t) \in \mathcal{S} \mid \exists t' \neq t,\, z \notin \Omega_{\mathcal S}(t') \}$ is non-empty then we refer to $\mathcal{S}$ as being invariant all time.

\subsection{Temporal Logic}
We define requirements for the intelligent intersection using temporal logic. This allows us to create high-level, human-readable requirements on how vehicles should be moving through the intersection. For example, in Fig.~\ref{fig:example}, we might specify the simple requirements ``vehicle should turn left and avoid collisions with other vehicles''. Using temporal logic, one can write a formal equation representing this requirement using operators that correspond to intuitive concepts in human language, i.e. ``always stay below 50 km/h'', ``always stay in lane'', ``eventually turn left'', or ``eventually exit intersection with 30 km/h'' (more examples listed in Fig.~\ref{fig:tlt}).
Others have leveraged the intuitive and rich specification capability of temporal logic in a variety of transportation problems~\cite{Wongpiromsarn2010, Tumova2016, Maierhofer2022}. In this work, we work with Linear Temporal Logic (LTL) since it yields the benefits of temporal logic, while being simple to work with and understand. Specifically, we use the operators $\{\ltlnot, \ltlor, \ltland, \ltluntil, \ltleventually, \ltlalways\}$, which correspond to the Boolean operators ``not'', ``or'', ``and'', and the temporal operators ``until'', ``eventually'', and ``always'', respectively. We note that similar methods to the one we present in this work can be applied to Signal Temporal Logic by adapting the work with the approach presented in~\cite{Yu2023}. We direct readers interested in seeing the same LTL syntax used in this work to~\cite{Jiang2024}.

\subsection{Temporal Logic Tree}
\begin{figure}[t]
    \centering
    \includegraphics[width=0.84\linewidth]{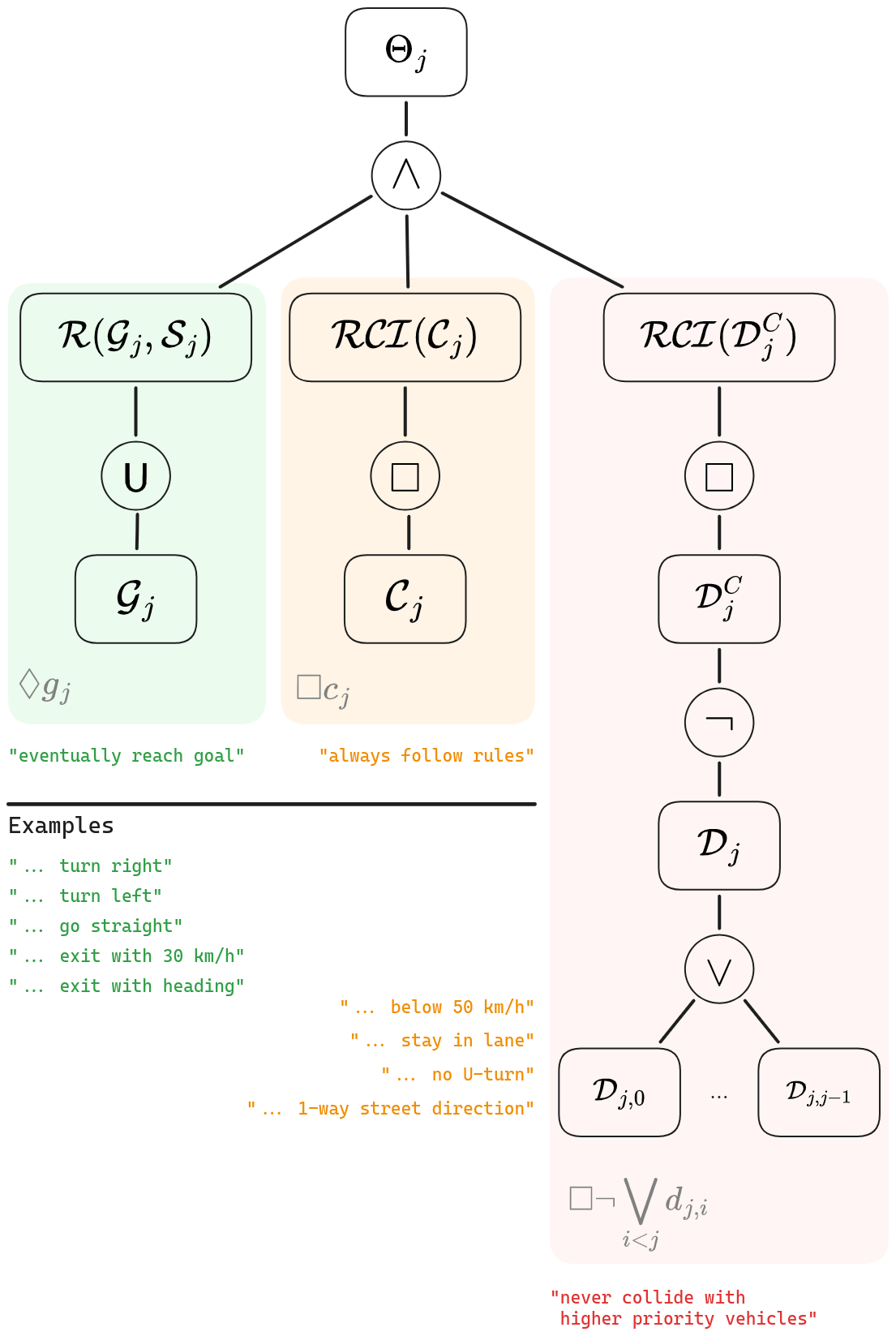}
    \caption{Illustration of the constructed TLT for the $j$th vehicle's individual LTL specification \eqref{eq:spp-ltl} that ensures safety at intersections.}
    \label{fig:tlt}
\end{figure}

Once we have specified an LTL specification for our multi-vehicle model, we need to check the feasibility of the specification. In this work, we will perform these computations using a computational model called Temporal Logic Trees~\cite{Gao_Abate_Jiang_Giacobbe_Xie_Johansson_2022} (example illustrated in Fig.~\ref{fig:tlt}).
Intuitively speaking, the leaf nodes of the temporal logic tree are the goals of the multi-vehicle system. From the goals, we perform a series of reachability analyses to find the joint set of feasible trajectories that satisfy the intersection specification. When the computation finishes, if the TLT has been successfully constructed, we know the intersection specification is feasible and possible to satisfy.
This verification result is detailed in~\cite[Theorem V.1]{Gao_Abate_Jiang_Giacobbe_Xie_Johansson_2022}. 
To construct temporal logic trees, we need to compute two kinds of reachable sets which are defined below.

\begin{definition}\label{def:brs}
\textbf{(Backward Reachable Tube)}
Given the full multi-vehicle system~\eqref{eq:multi_vehicle}, a computation time horizon $T$, a constraint time-state set $\mathcal C \subseteq \mathbb S \times \mathbb R$, and a target time-state set $\mathcal G \subseteq \mathbb S \times \mathbb R$, we define the backward reachable tube as
    \[\begin{split}
    \mathcal R(\mathcal G; \mathcal C) = \{ 
        (z, t) \; | \;
            & \exists u(\cdot) \in U, \: \\
            & \exists \tau \in [t, T), \: \zeta(\tau; z, t, u(\cdot)) \in \Omega_{\mathcal G}(\tau), \\
            & \forall \tau' \in [t, \tau), \:
            \zeta(\tau'; z, t, u(\cdot)) \in \Omega_{\mathcal C}(\tau')
    \},
    \end{split}\]
where $\mathcal R(\mathcal G; \mathcal C)$ contains the set of states that are able to reach the target set $\mathcal G$ while respecting the constraint set $\mathcal C$.
\end{definition}

\begin{definition}\label{def:rci}
\textbf{(Robust Control Invariant Set)} For the full multi-vehicle system~\eqref{eq:multi_vehicle}, computation time horizon $T$, and constraint set $\mathcal C \subseteq \mathbb{S}\times\mathbb{R}$, $\mathcal{RCI}(\mathcal C) \subseteq \mathbb{S}\times\mathbb{R}$ the largest robust control invariant set such that $\forall (z, t) \in \mathcal{RCI}(\mathcal C)$ there $\exists u(\cdot)\in U$ such that $\forall \tau \in [t, T]$, $\zeta(\tau; z, t, u(\cdot)) \in \mathcal C$.
\end{definition}

By using $\mathcal R(\cdot)$ and $\mathcal{RCI}(\cdot)$, we are able to fully construct temporal logic trees. For more details, we refer readers to~\cite{Gao_Abate_Jiang_Giacobbe_Xie_Johansson_2022}.

%% file: src/commands.tex
\pragmaonce

\newcommand{\BRS}{\text{BRS}}
\newcommand{\BRSmax}{\BRS^{max}}
\newcommand{\BRSmin}{\BRS^{min}}
\newcommand{\BRT}{\text{BRT}}
\newcommand{\BRTmaxany}{\BRT^{max}_{any}}
\newcommand{\BRTmaxall}{\BRT^{max}_{all}}
\newcommand{\BRTminany}{\BRT^{min}_{any}}
\newcommand{\BRTminall}{\BRT^{min}_{all}}

\newcommand{\ltltrue}{\textit{true}}
\newcommand{\ltlfalse}{\textit{false}}
\newcommand{\ltlnot}{\neg}
\newcommand{\ltlor}{\lor}
\newcommand{\Ltlor}{\bigvee}
\newcommand{\ltland}{\land}
\newcommand{\Ltland}{\bigwedge}
\newcommand{\ltlimply}{\rightarrow}
\newcommand{\ltlnext}{\bigcirc}
\newcommand{\ltluntil}{\,\mathsf{U}\,}
\newcommand{\ltlalways}{\square}
\newcommand{\ltleventually}{\lozenge}
\newcommand{\ltlsatisfy}{\models}

%% file: src/3_mot_ex.tex
\subsection{Problem Statement}\label{subsec:mot_ex}

In this work, we are interested in developing an approach for guaranteeing that vehicles are safely coordinated through intelligent intersections. For example, in Figure~\ref{fig:example}, we illustrate a T-intersection example that we will evaluate using our approach. The two vehicles are approaching the T-intersection, vehicle $\mathcal{V}_1$ from the right and vehicle $\mathcal{V}_2$ from below. Both intend to make a left turn, posing a potential risk of collision. For each of these vehicles, we can specify their overall behavior using the LTL formulae:
\begin{equation}
    \varphi_1 = \varphi_{\text{turn left}} \ltland \varphi_{\text{safety}},\ \varphi_2 = \varphi_{\text{turn left}} \ltland \varphi_{\text{safety}},
    \nonumber
\end{equation}
where $\varphi_1$ is $\mathcal V_1$'s specification and $\varphi_2$ is $\mathcal V_2$'s specification. These LTL formulae reflect each vehicle's individual specification of turning left while staying safe. Then, the specification for the intelligent intersection can be described by the following multi-vehicle LTL formula:
\begin{equation}\label{eq:example_formula}
    \varphi = \varphi_1 \ltland \varphi_2.
\end{equation}
This simple example is representative of the primary safety challenge of intersections: how can vehicles pass through the intersection while avoiding collisions? For the rest of the work, we will refer to this challenge as the ``intersection safety challenge.'' As was mentioned earlier, this challenge is addressed and solved by previous works. However, many of these solutions provide complete solutions that may be difficult to safely extend or adapt, as the safety guarantees are often built on the particular design decisions in the intersection management algorithm. Thus, much like~\cite{Maierhofer2022, Saraoglu2022}, we seek to leverage the richness of temporal logic to develop a safety framework that can solve for solutions to the intersection safety challenge in a way that can be easily adapted and built upon. Explicitly, given a multi-vehicle specification for an intersection, such as~\eqref{eq:example_formula}, we seek to automatically verify its feasibility
to guarantee the full specification is satisfied, while considering all the vehicle's dynamics and decision uncertainty.

%% file: src/4_method.tex
\section{Safety Verification for Intersections}\label{sec:method}

In this section, we develop our approach to finding solutions to the intersection safety challenge described in Section~\ref{subsec:mot_ex}. We start by posing the intersection safety challenge as a sequential path planning problem, inspired by the method developed in~\cite{Chen_Bansal_Tanabe_Tomlin_2017}. Then, we formalize the sequential path planning problem into LTL formulae. After we obtain LTL formulae, we detail how to use TLT to compute satisfaction sets for the sequential path planning specification. To make the approach more practical, we detail the computational approaches we employ to construct the TLT with a reasonable total computation time. Finally, we put everything together and detail the full verification approach that we can use and easily adapt to verify the safety of intersection rules.

\subsection{Sequential Path Planning for Intersections}

As the basis of our approach to solving the intersection safety challenge, we propose the formalization of the Sequential Path Planning (SPP) method, which is outlined in~\cite{Chen_Bansal_Tanabe_Tomlin_2017, Bansal_Chen_Tanabe_Tomlin_2021}. SPP is a structured approach for path planning in multi-vehicle scenarios. As suggested by the name, the key idea in SPP is to plan the paths of vehicles sequentially, prioritizing them based on a predefined order. In this method, when a higher priority vehicle $\mathcal{V}_i$ plans its path, it does so without considering subsequent vehicles. Since the path planning is sequential, when planning for a lower priority vehicle $\mathcal{V}_j$, where $i < j$, all admissible trajectories of $\mathcal{V}_i$ are already known. Consequently, $\mathcal{V}_i$ can be reserved in space and time, making it a known and deterministic obstacle for $\mathcal{V}_j$. The path planning problem for each vehicle $\mathcal{V}_j$ is then solved by computing the backward reachable set from a single-vehicle target set. After computing the backward reachable set, similar to Definition~\ref{def:brs}, we have a set that includes states from which $\mathcal{V}_j$ can reach its target within a specified time frame while avoiding all obstacles.
To make SPP more extensible and easily adaptable, we specify an LTL formulae that we will be able to leverage to create new intersection rules.

\subsection{LTL Specification of Sequential Path Planning} \label{sec:ltl-spec}

For specifying LTL formulae for SPP, we start by outlining the required specifications. First, we would like the vehicles to eventually reach their targets. Second, while they reach their targets, they should adhere to the traffic rules (speed limits, lane restrictions, etc.) in the intersection. Finally, while they reach their targets, they should also avoid colliding with other vehicles. As is done in SPP, instead of asking that the vehicles should avoid all other vehicles, we specify that it is enough that the vehicles only avoid the vehicles that are higher priority. 
We can write an LTL formula that collectively covers the listed specifications for an individual vehicle $\mathcal{V}_j$:
\begin{equation}\label{eq:spp-ltl}
	\varphi_j = \ltleventually g_j \ltland \ltlalways c_j \ltland \ltlalways \ltlnot \Ltlor_{i<j} d_{j,i}.
\end{equation}
Here, \(\ltleventually g_j\) corresponds to the requirement that $\mathcal V_j$ should eventually reach its goal, which is denoted by goal time-state set $\mathcal{G}_j\subseteq\mathbb{S}\times\mathbb{R}$. By including specific parts of $\mathcal V_j$'s state space in the goal set, a variety of goals can be represented, such as turning right, turning left, going straight, exiting the intersection with a specific speed, or exiting the intersection with a specific heading (as is listed in green in Fig.~\ref{fig:tlt}). Then, \(\ltlalways c_j\) corresponds to the requirement that $\mathcal V_j$ should follow traffic rules, which means its trajectories stays within a time-state constraint set $\mathcal{C}_j \subseteq \mathbb{S} \times \mathbb{R}$. Similarly to the goal set, a variety of traffic rules can be expressed through the constraint set, such as staying below the speed limit, staying in a specific lane while passing through the intersection, not allowing U-turns, and enforcing one-way street directions (as is listed in orange in Fig.~\ref{fig:tlt}). Finally, the last term, $\ltlalways \ltlnot \Ltlor_{i<j} d_{j,i}$, corresponds to the requirement that $\mathcal{V}_j$ avoids collisions with higher-priority vehicles. The proposition $d_{j,i}$ corresponds to the danger time-state set $\mathcal{D}_{j,i} \subseteq \mathbb{S} \times \mathbb{R}$, which are states where $\mathcal V_j$ is able to collide with a higher priority vehicle $\mathcal V_i$.
Notably, the highest priority vehicle $\mathcal{V}_1$ does not need to avoid any other vehicle. To handle this, we include a virtual vehicle $\mathcal{V}_0$ which cannot be collided with. Consequently, for any vehicle $\mathcal{V}_j$, $\mathcal{D}_{j,0} = \emptyset$ and $d_{j,0} = \textit{false}$. 

We note that the construction or computation of $\mathcal D_{j,i}$ is critically important to the efficiency of the intersection. When $\mathcal D_{j,i}$ is large,~\eqref{eq:spp-ltl} will result in $\mathcal V_j$ driving more conservatively and, in turn, less efficiently. To maximize the efficiency of the intersection, $\mathcal D_{j,i}$ should closely follow the true trajectory of $\mathcal V_i$. However, the less conservative $\mathcal D_{j,i}$ is, the higher the risk that $\mathcal V_i$ will accidentally leave $\mathcal D_{j,i}$. In other words, the computation of $\mathcal D_{j,i}$ introduces an important trade-off between safety and efficiency. The integration and development of computational approaches to computing $\mathcal D_{j,i}$ is not the focus of this paper and will be addressed in future work.

\subsection{Connecting LTL to Reachability Analysis}

We will now aim to bridge the gap between the LTL specification in \eqref{eq:spp-ltl} and the subsequent reachability analyses by constructing the TLT shown in Fig.~\ref{fig:tlt}. We keep in mind that the collective objective is to satisfy $\varphi = \Ltland_j \varphi_j$, yet the actual analyses will be made sequentially for each vehicle $\mathcal{V}_j$'s objective $\varphi_j$. 

\subsubsection{Computing Temporal Logic Trees}

As indicated by Fig.~\ref{fig:tlt}, the leaf nodes in the TLT represent the target proposition $g_j$, the state constraint proposition $c_j$ and the collision proposition $d_{j,i}$, respectively. Through the use of the target, state constraint, and collision propositions, we are able to freely encode and adapt different desired behaviors for the intersection. Once the specification is designed, the construction of the TLT proceeds by computing the reachable tubes $\mathcal R(\cdot)$ and $\mathcal{RCI}(\cdot)$ underlying the ``until'' ($\ltluntil$) and the ``always'' ($\ltlalways$) temporal operators, respectively. The ``eventually'' operator is a special case of the ``until'' operator and, thus, is also captured by computing $\mathcal R(\cdot)$. Then, the presence of Boolean operators ``not'' ($\ltlnot$), ``or'' ($\ltlor$), and ``and'' ($\ltland$) correspond to applying set complements, union, and intersection, respectively.

The application of the set operations underlying the Boolean operators is well-known and often exact. However, when an intersection is naively applied to two reachable tubes, this can result in an approximation error. This is due to the fact that the two reachable tubes may have targets or objectives that are conflicting and are not possible to simultaneously satisfy. This problem is sometimes known as the ``leaking corner problem''~\cite{Chen2018a, he2023efficient, Jiang2024}. We will refer to it as the ``conflicting objectives problem'' in this work. We address this conflicting objectives problem by recomputing the reachable tube of lower priority vehicles starting from the point in time where their tube intersects with the danger time-state set of higher priority vehicles. In the rest of this section, we detail the computation of the reachable tubes necessary for these SPP LTL formulae and the different computational techniques we use to reduce computational costs and avoid the conflicting objectives problem.

\subsubsection{Reachability Analysis for Intersections}
For intelligent intersection specifications, we construct temporal logic trees using HJ reachability analysis. The computational approach we use for HJ reachability analysis is a powerful approach that is based on finding the viscosity solution to a Hamilton-Jacobi-Isaacs Variational Inequality (HJI VI). For more details about this approach, we refer readers to~\cite{Chen_Tomlin_2018}). HJ reachability analysis is especially beneficial for addressing the intersection safety challenge due to it's ability to easily handle the nonlinear dynamics of vehicles and non-convex road geometries. Moreover, the resultant value functions from HJ reachability analysis can be used to efficiently compute the acceleration and steering rate limits for each vehicle~\cite{Jiang2024}.

One of the challenges for computing reachable tubes for intersections is the handling of timing in the intersection. In particular, many problems that reachable tubes are typically computed for are time invariant problems. For example, traffic rules and other state constraints, such as one-way street directions, do not typically depend on time.  However, the proposition $d_{j,i}$ is used to prevent collisions, so the corresponding $\mathcal{D}_{j,i}$ cannot be invariant since it must encode the movements of a higher priority vehicle. To address this, we adopt the double-obstacle HJI VI from~\cite{Fisac_Chen_Tomlin_Sastry_2015, Chen_Fisac_Sastry_Tomlin_2016}.
Specifically, consider $V_\mathcal{G}(z, t)$ and $V_\mathcal{C}(z, t)$ as two implicit surface functions representing the target and state constraints, respectively. Then the value function for $\mathcal{R}(\mathcal{G}; \mathcal{C})$ becomes:
\begin{equation}\label{eq:hji_vi}
        V(z, t) = \min_{\tau \in [t, T]}\max\{V_\mathcal{G}(z, \tau), \max_{\tau' \in [t, \tau]} V_\mathcal{C}(z, \tau') \}.
\end{equation}
We use~\eqref{eq:hji_vi} to compute $\mathcal{R}(\mathcal{G}; \mathcal{C}) = \{(z, t) | V(z, t) \leq 0\}$. When necessary, we compute the $\mathcal{RCI}(\cdot)$ in the same manner as is done in~\cite{Jiang2024}. However, as we explain in the remaining section, there are some cases where $\mathcal{RCI}(\cdot)$ does not need to be explicitly computed, yielding a reduction of computational cost for constructing the full TLT.

\begin{figure*}
    \centering
    \includegraphics[width=0.98\textwidth]{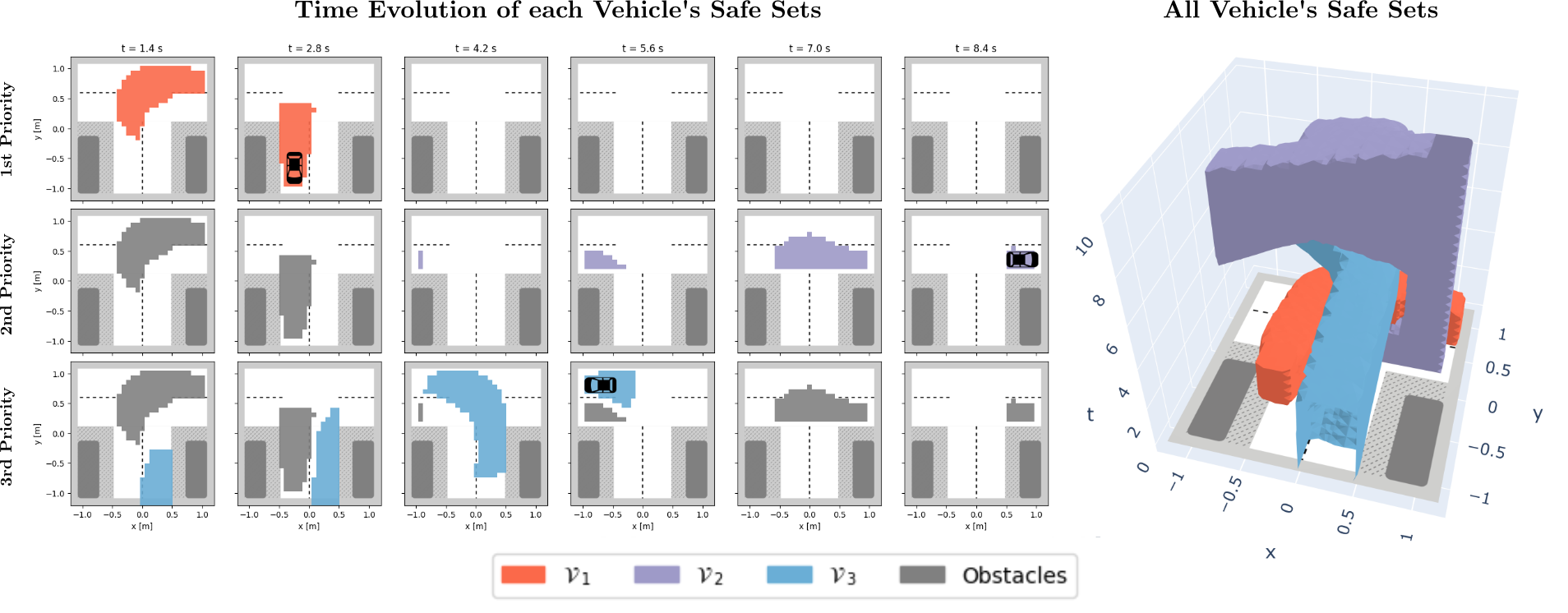}
    \vspace{-1ex}
    \caption{Shown are the safe sets for all vehicles over time (left part) and the sliced time snapshots of the safe sets for each vehicle, $\mathcal{V}_1$ (red), $\mathcal{V}_2$ (blue) and $\mathcal{V}_3$ (purple), with priority in the given order (right part). Since the safe sets are computed starting from the goal state, we mark the goal state of each vehicle with an icon. The top row of the right part shows the analysis done for $\mathcal{V}_1$ w.r.t its individual objective $\varphi_1$. Similarly, the second and third rows show the analyses done for $\mathcal{V}_2$ and $\mathcal{V}_3$, respectively, where the gray regions are the reachable sets of higher priority vehicles seen as obstacles.}
    \label{fig:simulation}
\end{figure*}

\subsection{Computational Approaches} \label{sec:comp_perf}

For the remainder of this section, we describe the different computational approaches we implement for reducing the computations needed for verifying an intersection specification and for avoiding unsafe approximations when the conflicting objectives problem emerges. Then, we end by describing the full computation we perform for checking the feasibility of the intersection specification. For particular details about the implementation of these computational approaches, we refer the reader the publicly available code.

\subsubsection{Simplifying the Reachability Analysis}
First, to reduce the reachability analysis necessary to construct the full TLT, we find that for our particular problem, we can compute one reachable tube for each vehicle in the multi-vehicle state space. A common case for intersections is that when vehicles reach their goals in the intersection, it is the same as leaving the intersection. In these cases, we can represent the satisfaction of $\ltleventually g_j \ltland \ltlalways c_j$ with the single reachable tube $\mathcal{R}(\mathcal{G}_j; \mathcal{C}_j)$. Normally, this reachable tube only represents the satisfaction of $c_j \ltluntil g_j$. However, since in the case where vehicles leave the intersection when they reach their goal, to satisfy $\ltleventually g_j$, we only need to keep track of trajectories that stop at $\mathcal{G}_j$ and do not need to worry about trajectories that will leave $\mathcal{G}_j$ afterwards. We find the same idea applies when including $\ltlalways\ltlnot\Ltlor_{i<j} d_{j,i}$ and that the full specification~\eqref{eq:spp-ltl} is verified by computing $\mathcal{R}(\mathcal{G}_j; \mathcal{C}_j \cap \mathcal{D}_j^C)$.

\subsubsection{Avoding the Conflicting Objectives Problem}
When higher priority vehicles are not present, then computing $\mathcal{R}(\mathcal{G}_j; \mathcal{C}_j)$ is sufficient to verify~\eqref{eq:spp-ltl}. If $\mathcal{G}_j$ and $\mathcal{C}_j$ are invariant, we can further improve performance by pre-computing the corresponding TLT subtrees for ${\ltleventually g_j \ltland \ltlalways c_j}$ in Fig.~\ref{fig:tlt} offline. However, when introducing vehicles to the intersection, we need to ensure that the conflicting objectives problem is avoided to fully ensure the safety of all vehicles. 
To do this, consider the time frame $T_{\mathcal{D}_{j}} \subseteq \mathbb{R}$ during which a vehicle $\mathcal V_j$ interact with higher priority vehicles and the conflicting objectives problem emerges. The time frame is a closed interval $T_{\mathcal{D}_{j}} = [t_a, t_b]$. 
For any $t > t_b$ it is sufficient to verify ${\ltleventually g_j \ltland \ltlalways c_j}$ since $\mathcal{D}_j$ will be empty and, consequently, $\ltlalways \ltlnot d_{j,i}$ will be true. Assuming the target and state constraints are invariant, then this can be done offline. On the other hand, for any $t \le t_b$ we need to recompute the analysis with the added collision constraint that ensures a safe interaction with higher priority vehicles. By doing this, we reduce the total amount of reachability analysis to performed online and avoid the conflicting objectives problem.

\subsubsection{Full Computation}
Finally, our approach to computing the TLT in Figure~\ref{fig:tlt} is the following. We start by computing $\mathcal{R}(\mathcal{G}_j; \mathcal{C}_j)$ for vehicle $\mathcal{V}_j$ offline and store it in memory. We call this step the ``Offline Pass''. Then,
we perform a second reachability analysis online, which we call the ``Online Pass''. Specifically,  the online pass updates the solution of $\mathcal{R}(\mathcal{G}_j; \mathcal{C}_j)$ over $t \leq t_b$ with
\begin{equation}\begin{split}
\mathcal{R}(
    &\mathcal{G}_j \cup \{(z,t_b) \mid z \in \Omega_{\mathcal{R}(\mathcal{G}_j, \mathcal{C}_j)}(t_b)\};
    \; \mathcal{C}_j \cap \mathcal{D}_j^C).
\end{split}\end{equation}
Incorporating these two passes enables us to efficiently check the satisfaction of the intersection specification.

In summary, by using this approach, we develop a verification method where we can freely design and adapt the requirements on the behavior of the vehicles passing through the intersection using the expressiveness of temporal logic. Then, by constructing temporal logic trees using HJ reachability analysis, we can handle general nonlinear dynamics and complex constraints in the state space of the vehicle model. In the next section, we evaluate the practicalities of applying this method to a T-intersection example.

%% file: src/5_experiments.tex
\section{Numerical Results}\label{sec:exp}

In this section the T-intersection scenario, as described in Section \ref{subsec:mot_ex}, is presented in simulation. The simulation, shown in Fig. \ref{fig:simulation}, includes three vehicles that cross the intersection in a way that risk collision if there is no coordination between them. Each vehicle is modelled by~\eqref{eq:single_vehicle} with working space and control constraints: $\mathbb{S}_i = \{ z_i \in \mathbb{R}^5 \mid -1.2 \leq x_i \leq 1.2, -1.2 \leq y_i \leq 1.2, -\pi \leq \theta_i \leq \pi, -\pi/5 \leq \delta_i \leq \pi/5, 0 \leq v_i \leq 1  \}, \ \mathbb{U}_i = \{ u_i \in \mathbb{R}^2 \mid -\pi \leq s_i \leq \pi, -0.5 \leq a_i \leq 0.5\}.$
The roads enforce constraints on the vehicles' heading, except in the middle of the intersection. Furthermore, a lower-bound speed limit of 0.4 m/s is set to prevent the vehicles from stopping and blocking the way for following vehicles. Finally, for each road we define entry and exit targets. For example, $\mathcal{V}_2$ will enter at $\mathbb{G}_2^\textit{enter} = \{ z_2 \in \mathbb{S}_2 \mid 0 \le x_2 \le 0.5, -1.2 \le y_2 -0.7\}$.
The reachability analysis is performed using the Python package \verb|hj_reachability|\footnote{https://github.com/StanfordASL/hj\_reachability} which solves the HJI VI on a $31 \times 31 \times 31 \times 7 \times 11$ grid of the discretized single-vehicle state space $\mathbb{S}_i$. \verb|hj_reachability| can compute this on the GPU. In Table \ref{tab:comp_time} we show how long these operations take on an NVIDIA GeForce RTX 2080 Ti. 

In Figure \ref{fig:simulation}, we show the time evolution of the computed safe sets of each vehicle. The  analysis' time horizon starts at $t=0$, which is the current time, and ends at $t=10$. The top row shows the evolution of the highest priority vehicle. The plotted safe set corresponds to all of the locations the highest priority vehicle can be in to satisfy the intersection specification. Then, after the safe set of the highest priority vehicle is computed, the safe set is passed to the computation of the next vehicle to be used as a danger set (grey set). We repeat this for each lower priority vehicle. In other words, the red and purple sets are the danger sets ($\mathcal{D}_{3,1}$ and $\mathcal{D}_{3,2}$) for the 2nd and 3rd priority vehicles, respectively.
Interestingly, we see the reachable set $\Theta_3$ as computed in the online pass for $\mathcal{V}_3$ in blue. Here, we clearly see how the reachability analysis ensures that $\mathcal{V}_3$ avoid the higher priority vehicles. The online pass is computed over the time frame $T_{\mathcal{D}_3} = [0, 6]$ during which it removes states that would otherwise lead to a collision with either $\mathcal{V}_1$ or $\mathcal{V}_2$. 

\begin{table}[t]
    \centering
    \begin{tabular}{c|cc}
    \textbf{Vehicle}& \textbf{Offline Pass} [s]&\textbf{Online Pass} [s]\\
    \hline
    $\mathcal{V}_1$& 16.11&0.0\\
    $\mathcal{V}_2$&  13.68& 6.71\\
    $\mathcal{V}_3$&  13.17& 8.25\\
    \end{tabular}
    \caption{Computational time of the reachability analyses for each vehicle on NVIDIA GeForce RTX 2080 Ti.}
    \label{tab:comp_time}
\end{table}

These results provide preliminary indication that the presented method can be used in practical settings. Although the computation times reported in Table~\ref{tab:comp_time} are not fast enough to be used in cases where the verification should occur while vehicles are in the intersection, the computations are fast enough if the vehicle has not arrived to the intersection yet. Moreover, in this work, we do not explore a variety of computational techniques that can further optimize the computation time for verifying the intersection safety. For example, in the case where a vehicle needs to be rescheduled, the tubes illustrated on the right-side of Fig.~\ref{fig:simulation} could simply be moved up or down the time-axis at almost no computational cost. Furthermore, due to the richness of the value function underlying the safe sets, the valid acceleration and steering rates the vehicles should implement can also be computed with a very low computational complexity~\cite{Jiang2024}.

%% file: src/6_conclusion.tex
\section{Conclusion}\label{sec:conc}

In this work, we propose a framework based on LTL and Hamilton-Jacobi reachability analysis for ensuring the safety at intelligent intersections. By formalizing SPP into LTL formulae, we leverage temporal logic trees to break down the intersection safety problem into a series of Hamilton-Jacobi reachability analyses. Due to this approach, the safety framework is able to handle changes in the intersection specification, while maintaining safety guarantees. We illustrate the framework's utility on a simulated T-intersection example, where we show we are able to verify that the vehicles can pass through the intersection safely. While we include several optimizations in our implementation to reduce the total computational times in the T-intersection example, an important future work will be to further reduce the computation time by employing techniques for directly addressing the conflicting objectives problem, such as the technique presented in~\cite{he2023efficient}. Moreover, we are building on the theoretical foundation of the presented framework to design and implement an intelligent intersection management system that integrates scheduling and V2I communication, so we can use the testbed presented in~\cite{MunhozArfvidsson2024} to evaluate the framework's performance with real hardware and communication networks in the loop.